\documentstyle[12pt]{article}
  \advance\oddsidemargin  by -1in
  \advance\evensidemargin by -1in
  \advance\topmargin      by -0.5in
\makeindex

\def\Pom{{\bf I\!P}}
 
\def\lsim{\mathrel{\rlap{\lower4pt\hbox{\hskip1pt$\sim$}}
    \raise1pt\hbox{$<$}}}         
\def\gsim{\mathrel{\rlap{\lower4pt\hbox{\hskip1pt$\sim$}}
    \raise1pt\hbox{$>$}}}         
\def\beq{\begin{equation}}
\def\eeq{\end{equation}}
\def\bea{\begin{eqnarray}}
\def\eea{\end{eqnarray}}

\begin{document}

\begin{flushright}
{\em
FZ-IKP(TH)-2000-30}
\end{flushright}

\vspace{1.0cm}
\begin{center}
{\Large \bf Diffractive
 beauty photoproduction as a short distance probe of QCD pomeron
\vspace{1.0cm}}

{\large \bf   V. R. Zoller}\\
\vspace{0.5cm}
{\em Institute for  Theoretical and Experimental Physics,
Moscow 117218, Russia\\
E-mail: zoller@heron.itep.ru} \vspace{0.5cm}\\
{\bf Abstract}
\end{center}
High-energy open beauty
photoproduction probes the vacuum exchange at  distances $\sim 1/m_b$
and detects significant  corrections to the 
BFKL asymptotics coming from the  subleading  vacuum poles.
 We show that the
interplay of leading and subleading vacuum exchanges gives rise to the
cross section $\sigma^{b\bar b}(W)$ growing much faster than it is
prescribed by the exchange of the leading pomeron trajectory with
intercept $\alpha_{\Pom}(0)-1=\Delta_{\Pom}=0.4$. Our calculations
within the  color dipole BFKL model are in
agreement with the recent determination of $\sigma^{b\bar b}(W)$ by the H1
collaboration.  The comparative analysis
 of diffractive photoproduction of  beauty, charm and light quarks exhibits
the hierarchy of pre-asymptotic pomeron intercepts  which follows the
hierarchy of corresponding hardness scales.
  We comment  on the phenomenon of decoupling of soft and
subleading BFKL singularities at the scale of elastic $\Upsilon(1S)$
 -photoproduction which results in precocious color dipole
 BFKL asymptotics of the
process $\gamma p \to \Upsilon p$.

\vspace{0.75cm}


 In this communication  we address the issue of  open beauty photo- and
electroproduction
\beq
\gamma^*p\to b\bar b X
\label{eq:GPBBX}
\eeq
at  large values of  Regge parameter
\beq
x^{-1}={W^2+Q^2\over 4m^2_b+Q^2}\gg 1\,,
\label{eq:X}
\eeq
where $W$ is the c.m.s. collision energy, $Q^2$ is the photon virtuality and
the mass of beauty quark $m_b$ sets the  natural scale of the process
(\ref{eq:GPBBX}).

In color dipole (CD) approach the  excitation of heavy flavors
 at small-$x$
is described in terms of interaction of small size
 quark-antiquark $b\bar{b}$ color
dipoles in the photon.
This makes the reaction (\ref{eq:GPBBX})
a sensitive probe of
short distance properties of vacuum exchange in QCD. The
interaction of color dipole ${\bf r}$  with
the target proton
 is described by the beam, target and flavor  independent
color dipole-proton cross section $\sigma(x,{\bf r})$. The
contribution of excitation of open beauty to photo-absorption cross
section is given by color dipole factorization formula 
\cite{GUNION,NZ91,MUELLER}
\beq
\sigma^{b\bar b}(x,Q^2)=
\int_0^1 dz\int
d^2{\bf r}\left[|\Psi_L^{b\bar b}(z,{\bf r})|^2
+|\Psi_T^{b\bar b}(z,{\bf r})|^2\right]\sigma(x,{\bf r}) \,,
\label{eq:2.4}
\eeq
 where $|\Psi_{L,T}^{~b\bar{b}}(z,{\bf{r}})|^{2}$ is
a probability to find in the photon the $b\bar{b}$ color dipole with the
beauty quark carrying fraction $z$ of the photon's light-cone momentum.
The well known result of \cite{NZ91} for the transverse (T) and longitudinal
(L)  photons is
\beq
|\Psi_T^{b\bar b}(z,r)|^2={\alpha_{em}\over 6\pi^2}
\left\{\left[z^2+(1-z)^2\right]\varepsilon^2K_1(\varepsilon r)^2
+m_b^2K_0(\varepsilon r)^2\right\},
\label{eq:2.2}
\eeq
\beq
|\Psi_L^{b\bar b}(z,r)|^2={2\alpha_{em}\over 3\pi^2}
Q^2z^2(1-z)^2K^2_0(\varepsilon r)\,,
\label{eq:2.3}
\eeq
where $K_{0,1}(y)$ are the modified Bessel functions, 
 $\varepsilon^2=z(1-z)Q^2+m_b^2$
and $m_b=4.7\,{\rm GeV}$ is the $b$-quark mass.
 Here we focus on the open beauty
photoproduction cross section.
   At $Q^2\ll m^2_b$ the eq.(\ref{eq:2.4})
takes the form

\beq
\sigma^{b\bar b}(x)\simeq {\alpha_{em}\over 6\pi}\int dr^2 m_b^2
\left[{2\over 3}K_1(m_b r)^2+K_0(m_b r)^2  \right]\sigma(x,r)
\label{eq:2.5}
\eeq

Because for small dipoles   $\sigma(x,r)\propto r^{2}$,
the dipole size integration in (\ref{eq:2.5})
is well convergent at small $r$
 and the integrand  (\ref{eq:2.5})
  has a peak at $ r \simeq 1/m_b$. At very high $Q^{2}\gsim 4m^2_b$ the
peak develops a plateau for dipole sizes in the interval
 $(m_b^2+Q^2/4)^{-1} \lsim r^2\lsim m_b^{-2}$.
So, for moderate photon virtualities $Q^{2}\ll 4m^2_b$ excitation of open
 beauty probes (scans) the dipole cross section at a special dipole
 size $r_{S}$, $r_{S}\sim {1/m_{b}}$.


The CD  cross section $\sigma(x,r)$
 at  $ x\ll 1$
satisfies the CD BFKL equation
\beq
{\partial \sigma(x,r) \over \partial \log(1/x)}={\cal K}\otimes \sigma(x,r)\,,
\label{eq:3.1}
\eeq
with the kernel ${\cal K}$  \cite{PISMA1}.
The basis of
 the CD BFKL-Regge expansion for
 $\sigma(x,r)$ \cite{NZZ94,PION}
\beq
\sigma(x, r)=\sum_n \sigma_n(r)\left({x_0\over x}\right)^{\Delta_n}.
\label{eq:3.5}
\eeq
form the solutions of the eigen-value problem
\beq
{\cal K}\otimes \sigma_{n}=\Delta_{n}\sigma_{n}(r)\,
\label{eq:3.2}
\eeq
 with Regge behavior,
$\sigma_{n}(x,r)=
\sigma_{n}(r)\left({x_0/ x}\right)^{\Delta_{n}}.$

The properties of the  CD BFKL equation and the choice of
physics motivated boundary condition at $x_0=0.03$ were discussed
in detail elsewhere
\cite{NZDelta,NZHERA,NZZ97,DER,NZcharm},
here we only recapitulate features
relevant to the considered problem.

The leading eigen-function $\sigma_0(r)$
for ground state i.e., for the rightmost hard BFKL pole, is node free.
The  subleading  eigen-function for excited state $\sigma_n(r)$ has $n$ nodes.
We find $\sigma_n(r)$ numerically \cite{NZZ97,DER}, for the semi-classical
analysis see Lipatov \cite{Lipatov}. The intercepts  follow
to a good approximation the law $\Delta_{n}= \Delta_{0}/(n+1)$.
 Within our specific
infrared regularization
we find $\Delta_{0}=0.402$, 
$\Delta_{1}=0.220 $ and $\Delta_{2}=0.148$ \cite{NZZ97}.
The  node of $\sigma_{1}(r)$ is located at $r=r_1\simeq
0.056\,{\rm fm}$ (see Fig.1), for larger $n$ the rightmost node moves to a
 somewhat larger $r=r_1\sim 0.1\, {\rm fm}$.
As we discussed elsewhere \cite{NZZ97}, for still higher solutions,
 $n\geq 3$,
all intercepts are very small anyway, $\Delta_n\ll \Delta_{0}$,
 For
this reason, for the purposes of practical phenomenology we can truncate
expansion (\ref{eq:3.6}) at $n=3$ lumping in the term $n=3$ contributions
of still higher singularities with $n\geq 3$ (see Fig.1). The term $n=3$
 is  endowed
 with the effective intercept $\Delta_3=0.06$ \cite{NZcharm2}.
 The  truncated expansion
reproduces the numerical
solution $\sigma(x,r)$ 
of CD BFKL equation (\ref{eq:3.1})
in the wide range of dipole sizes $10^{-3}\lsim r \lsim 10$ fm with accuracy 
 $\simeq 10\%$ even at moderately small  $x$.

 Each  CD BFKL eigen-cross section $\sigma_n$ via
 equation 
\beq
f_n^{b}(Q^2)={Q^2\over 4\pi^2\alpha_{em}}
\int_0^1 dz\int
d^2{\bf r}\left[|\Psi_L^{b\bar b}(z,{\bf r})|^2+|\Psi_T^{b\bar
b}(z,{\bf r})|^2\right]\sigma_n(r)                      \,.
\label{eq:fm}
\eeq
 defines the corresponding
eigen-structure function (SF)  $f_n^{b}(Q^2)$ and we arrive at the 
CD BFKL-Regge
expansion for the beauty SF of the proton
$(n=0,1,2,3,{\rm soft})$
\beq
F_2^{b}(x,Q^2)={Q^2\over 4\pi^2\alpha_{em}}\sigma^{b\bar b}(x,Q^2)
 =\sum_{n} f_n^{b}(Q^2)\left(x_0\over x\right)^{\Delta_n}\,.
\label{eq:3.6}
\eeq
Analytical parameterization for eigen-SF $f_n^{b}(Q^2)$ is presented in
Appendix. 
Notice that numerically
$r_{1} > r_{S}$
and  the rightmost nodes of subleading eigen-cross sections
 $\sigma_n(r)$
are located to the right of the peak of the integrand (\ref{eq:2.5})
(see Fig.1).
Consequently, in the calculation of open beauty eigen-SFs $f_n^{b}(Q^2)$
one scans the eigen-cross section in between of the first and the second
node. Hence,  subleading $f_n^{b}(Q^2)$  which are negative valued
 in a wide range of $Q^2$. This
point is illustrated in Fig.~2 in which the subleading BFKL-to-rightmost
BFKL and soft-pomeron-to-rightmost BFKL ratio  of eigen-SFs
$r_n(Q^2)=f^b_n(Q^2)/f^b_0(Q^2)$ is shown.
  Because a probability to find large color dipoles in the photon
decreases rapidly with the quark mass, the contribution from
 energy-independent soft-pomeron
exchange to open beauty excitation is very small down to $Q^{2}=0$.


We comment first on the results on $\sigma^{b\bar b}(W)$.
 Because the CD BFKL-Regge
expansion for color dipole-dipole cross section has already been fixed
from the related and highly successful phenomenology of light flavor and charm
contributions to the proton SF \cite{PION,NZZ97,DER,NZcharm} the CD BFKL
predictions for the beauty SF of the proton are parameter free.

 In Fig.3
our evaluation of the $W$-dependence of open charm (upper solid curve) and
open beauty photoproduction cross sections (lower solid curve)
 is confronted to the HERA data \cite{DATAHERACC,H1BEAUTY}. 
The H1 analysis of the data on $\sigma(\gamma p\to b\bar b X)\equiv 
\sigma^{b\bar b} $ for the
collision energy in the range $94<W<266$ GeV with $\langle W\rangle\simeq 
180$ GeV
gives \cite{H1BEAUTY}
$$\sigma^{b\bar b}=111\pm10^{+16}_{-11}\pm 17\,nb.$$
 The CD BFKL-Regge approach
 results in (Fig.3)
$$\sigma^{b\bar b}\simeq 81\,nb$$
at $ W \simeq 180$ GeV.
 For an alternative interpretation of beauty photoproduction
see \cite{FRIXIONE}. We differ from \cite{FRIXIONE} in estimate of the effect
of $\log(1/x)$-evolution. The latter seems to be  strongly
 underestimated in \cite{FRIXIONE}. 

As we have emphasized above, the characteristic feature of the QCD pomeron
 dynamics at distances $\sim m_b^{-1}$ is large negative valued 
 contribution to
$\sigma^{b\bar b}$ coming from subleading BFKL singularities. 
 Consequences of this observation
 for the exponent of the
 energy dependence of the cross section
\beq
\sigma^{b\bar b}(W)\propto W^{2\Delta_{\rm eff}}
\label{eq:SBBEXP}
\eeq
 are quite interesting.
In terms of the ratio
$r_n(Q^2)$  (Fig.2) taken at $Q^2\ll 4m^2_b$
and denoted by $r_n(0)$ the exponent
$\Delta_{\rm eff}$ reads (n=1,2,3,soft)
\beq
\Delta_{\rm eff}=\Delta_{0}{1+\Delta^{-1}_0\sum_n \Delta_n 
r_n(0) ({x_0/ x})^{\Delta_n-\Delta_0}\over
1+\sum_n r_n(0) ({x_0/ x})^{\Delta_n-\Delta_0} }
\label{eq:DELTAEFF}
\eeq
Because all coefficients $r_n(0)$
 in eq.(\ref{eq:DELTAEFF})
are negative, except $r_{\rm soft}(0)>0$, at HERA energies
 the effective
 intercept overshoots the asymptotic value
 $\Delta_{\Pom}\equiv \Delta_0=0.402$
 (see Fig.4).
 At still higher collision energies  both  the soft and subleading hard BFKL
  exchanges become rapidly Regge suppressed. This results in
 decreasing $\Delta_{\rm eff}$ down to
 $\Delta_{\Pom}$.

For comparison,
 in photoproduction of open charm which scans the color dipole cross section
 at distances
$\sim 1/m_c$, in the vicinity of the rightmost node (see Fig.1),
 there is a strong cancellation between soft and  subleading
contributions to $\sigma^{c\bar c}(W)$ \cite{NZcharm,NZcharm2}.
 Consequently, for this dynamical
 reason in open  charm photoproduction we have an effective one-pole
picture and the effective pomeron intercept
 $\Delta_{\rm eff}\simeq \Delta_{\Pom}$.

In photoproduction of light flavors the CD cross section $\sigma_n(r)$ 
is close to the
 saturation regime $\sigma_n(r)\propto const$
and all subleading and soft terms of the CD BFKL-Regge expansion
 are positive valued
and numerically important (see \cite{PION} for more details).
This is the dynamical reason for smallness of a pre-asymptotic
 pomeron intercept in  photoproduction of light flavors (see Fig.4).
Notice that it must not be taken at face value 
for $W\sim 1$ TeV because of likely strong absorption
corrections. However,   the  hierarchy of  pre-asymptotic 
intercepts which
brings to light the internal dynamics of leading-subleading cancellations at
different hardness scales  
 should withstand unitarity effects.

In Fig.5  we presented our predictions for the beauty structure functions
in DIS on protons as a function of the Bjorken variable 
$x_{Bj}=Q^2/(W^2+Q^2)$. The solid curve corresponds to the
complete expansion (\ref{eq:3.6}) while 
 the long-dashed
curve is the leading hard pole approximation, $F_2^{b}(x,Q^2)\simeq
 f_0^{b}(Q^2)\left(x_0/ x\right)^{\Delta_0}$.
 In agreement with the nodal structure of subleading
eigen-SFs the latter  over-predicts
 $F_2^b$ significantly because the negative valued contribution from
 subleading hard BFKL
 exchanges
overtakes the soft-pomeron exchange, see Fig.~2, and the background from
subleading hard BFKL exchanges is substantial for all $Q^{2}$.
 We do not stretch the theoretical
curves  to  $x> x_{0}=0.03$ beyond the applicability region of CD
 BFKL-Regge expansion (\ref{eq:3.6}).

Recently the cross section of elastic $\Upsilon(1S)$ meson
photoproduction has been measured at HERA \cite{ELBEAUTY}. Quarks in
 $\Upsilon$ meson are nonrelativistic and $|\gamma\rangle\propto
 m_bK_0(m_br)$. The forward $\gamma\to \Upsilon$ transition matrix element
 $\langle\Upsilon|\sigma_n(r)|\gamma\rangle$ is controlled by the
 product $\sigma_0(r)K_0(m_br)$ \cite{SCAN}
and the amplitude of  elastic
 of $\Upsilon(1S)$ photoproduction
  is dominated by the contribution from the dipole
sizes $r\sim r_{\Upsilon}=A/m_{\Upsilon}$
with $A= 5$. The crucial observation   is that at distances 
$r\sim r_{\Upsilon}$
  cancellation between soft and subleading contributions to the elastic
photoproduction cross section  results in the exponent
  $\Delta$ in 
\beq
{d\sigma(\gamma p\to\Upsilon p )\over dt}|_{t=0}\propto W^{4\Delta}
\label{eq:ELBB}
\eeq
 which is very close to $\Delta_{\Pom}$,  $\Delta=0.38$ \cite{JETPVM}. This
observation  appears  to be in
agreement with the cross section rise observed by ZEUS\&H1 
\cite{ELBEAUTY}.

{\bf Appendix.} For the practical  applications it is convenient to
have analytical parameterization for eigen-SFs
$f^{b}_n(Q^2)$, which for the rightmost hard BFKL pole $n=0$ is of the form
\beq
f^b_0(Q^2)=
{aR^2Q^2\over{1+ R^2Q^2 }}
\left[1+c\log(1+r^2Q^2)\right]^{\gamma},\,\,\gamma={4\over 3\Delta_0}
\label{eq:F20C}
\eeq
where $a=3.877\cdot 10^{-2}$, $R^2=1.036\cdot 10^{-2}$ GeV$^{-2}$,
 $c=0.1482$, 
$r^2=1.004\cdot 10^{-2}$ GeV$^{-2}$.
 For the subleading  pole
with $n=1$ the eigen-SF  $f^b_1(Q^2)$ is of the form (\ref{eq:F20C})
 with $a=-4.684\cdot 10^{-3}$,
 $R^2=1.352\cdot 10^{-2}$ GeV$^{-2}$, $c=0.2002$, 
$r^2=1.400\cdot 10^{-2}$ GeV$^{-2}$,
 $\gamma=4\delta/3\Delta_0$ and $\delta=1.621$.
 For still higher $n$ we have
\beq
f^b_2(Q^2)=af_0(Q^2)\left(1-z/z_1\right)\left(1+z/z_2\right)
/\left(1+z/z_3\right)\,,
\label{eq:FN2}
\eeq
where
$z=\log(1+r^2Q^2)^{\gamma}$,
$\gamma=1.935$, $r^2=5.727\cdot 10^{-2}$ GeV$^{-2}$,
 $a=-0.153$, $z_1=47.46$, $z_2=11.56$, $z_3=4944.7$
and
\beq
f^b_3(Q^2)=af_0(Q^2)\left(1-z/z_1\right)\left(1-z/z_2\right),
\label{eq:FN3}
\eeq
where
$z=\left[1+c\log(1+r^2Q^2)\right]^{\gamma}-1,$
$\gamma={4\delta/ 3\Delta_0}$, 
$\delta= 1.362$, $c=4.92\cdot 10^{-2}$,
$r^2=5.002\cdot 10^{-3}$ GeV$^{-2}$, 
$a=- 0.3386$, $z_1=0.6844$ and $z_2=3.03$.
The soft component of the beauty SF with $\Delta_{\rm soft}=0$
was derived from eq.(\ref{eq:2.4}) with
 $\sigma_{\rm soft}(r)$ taken from
\cite{JETPVM}.


{\bf Acknowledgments: } The author is grateful to J.Speth for hospitality
 at IKP(Theory) FZ-Juelich.
 This work was supported partly by the grant
 INTAS-97-30494.



\newpage

{\bf Figure captions}
\begin{enumerate}
\item[{\bf Fig.1}]
The CD BFKL eigen functions plotted as $\sigma_n(r)/r$ for n=0,1,2.
 The background  term, $n=3$, which
 is a combination
of higher CD BFKL solutions with $n\geq 3$ is also shown.

\item[{\bf Fig.2}]
The subleading hard-to-rightmost hard and soft-pomeron-to-rightmost hard
ratio of eigen-structure functions
 $r_n=f_n^{b}(Q^2)/f_0^{b}(Q^2)$  as a function
$Q^{2}$ .

\item[{\bf Fig.3}]
Predictions from CD BFKL-Regge factorization for the heavy flavor $Q\bar Q$
 photoproduction
cross section.
 The lower  curve is a result
of the complete
CD BFKL-Regge expansion for $\sigma^{b\bar b}(W)$. The upper 
 curve corresponds to
  $\sigma^{c\bar c}(W)$ calculated 
 with $m_c=1.3$ GeV. The data points are from H1\&ZEUS HERA experiments
 \cite{DATAHERACC, H1BEAUTY}.

\item[{\bf Fig.4}]
The W-dependence of the  exponent $\Delta_{\rm eff}$ in parameterization 
of the photoproduction
 cross section $\sigma^{q\bar q}(W)\propto W^{2\Delta_{\rm eff}}$ for
 different quark flavors. 

\item[{\bf Fig.5}]
Prediction from CD BFKL-Regge factorization for the beauty structure function
of the proton $F^{b}_2(x,Q^2)$ as a function of
the Bjorken variable $x_{Bj}$.

\end{enumerate}

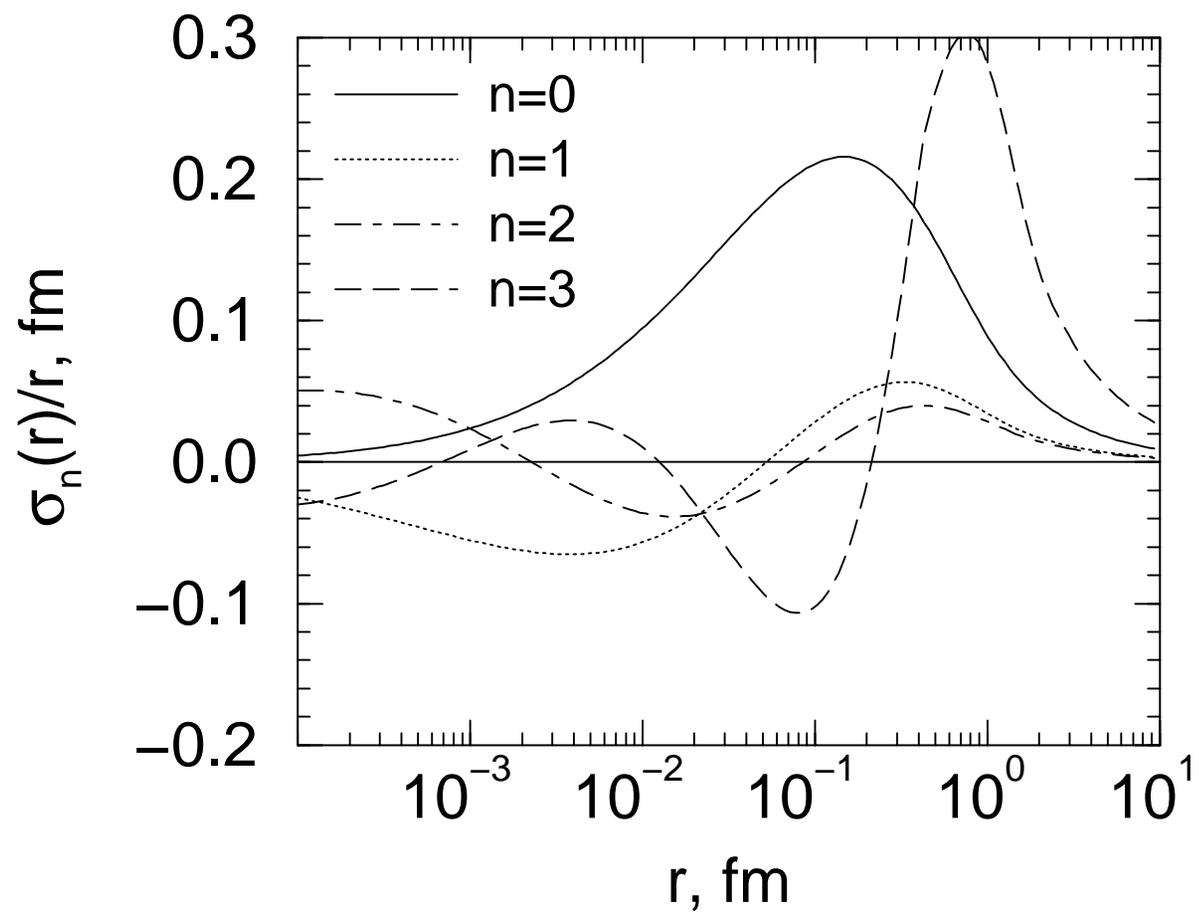
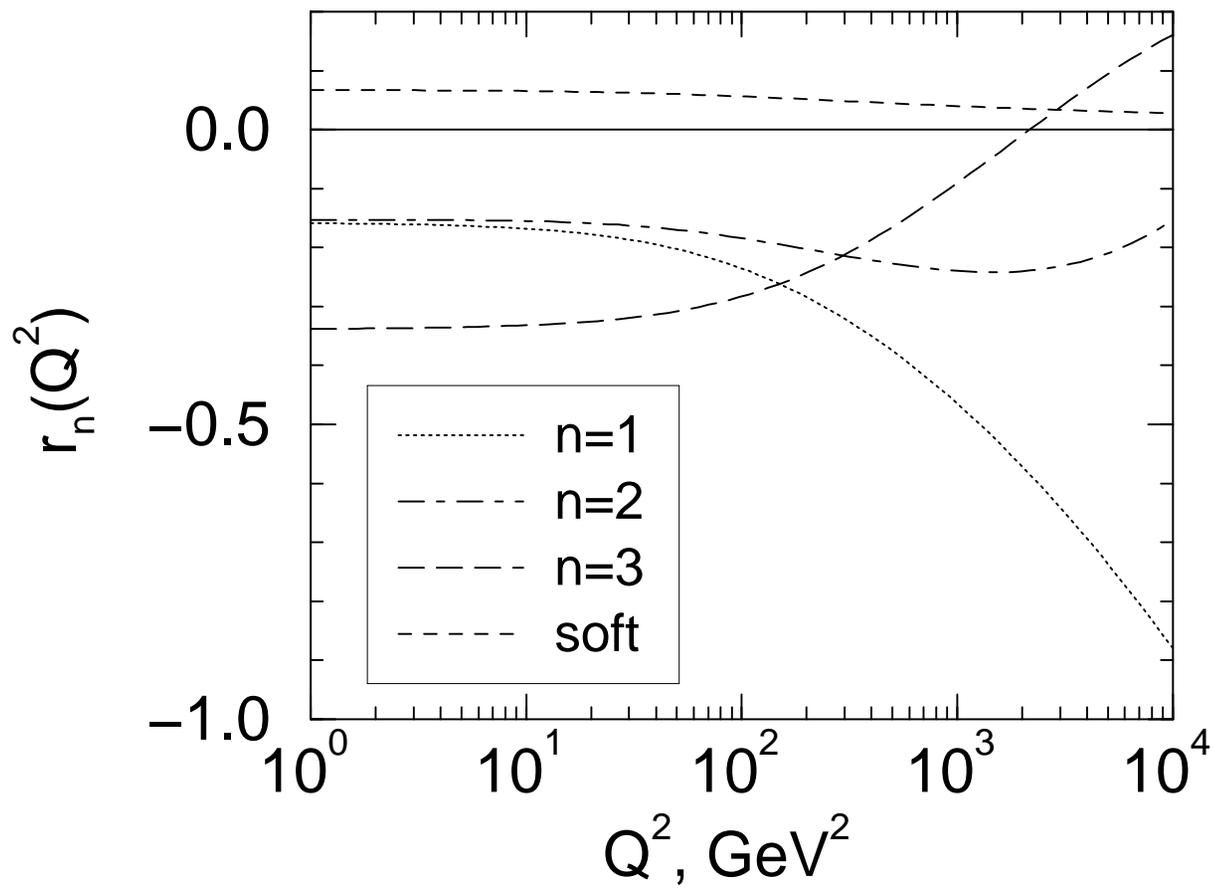
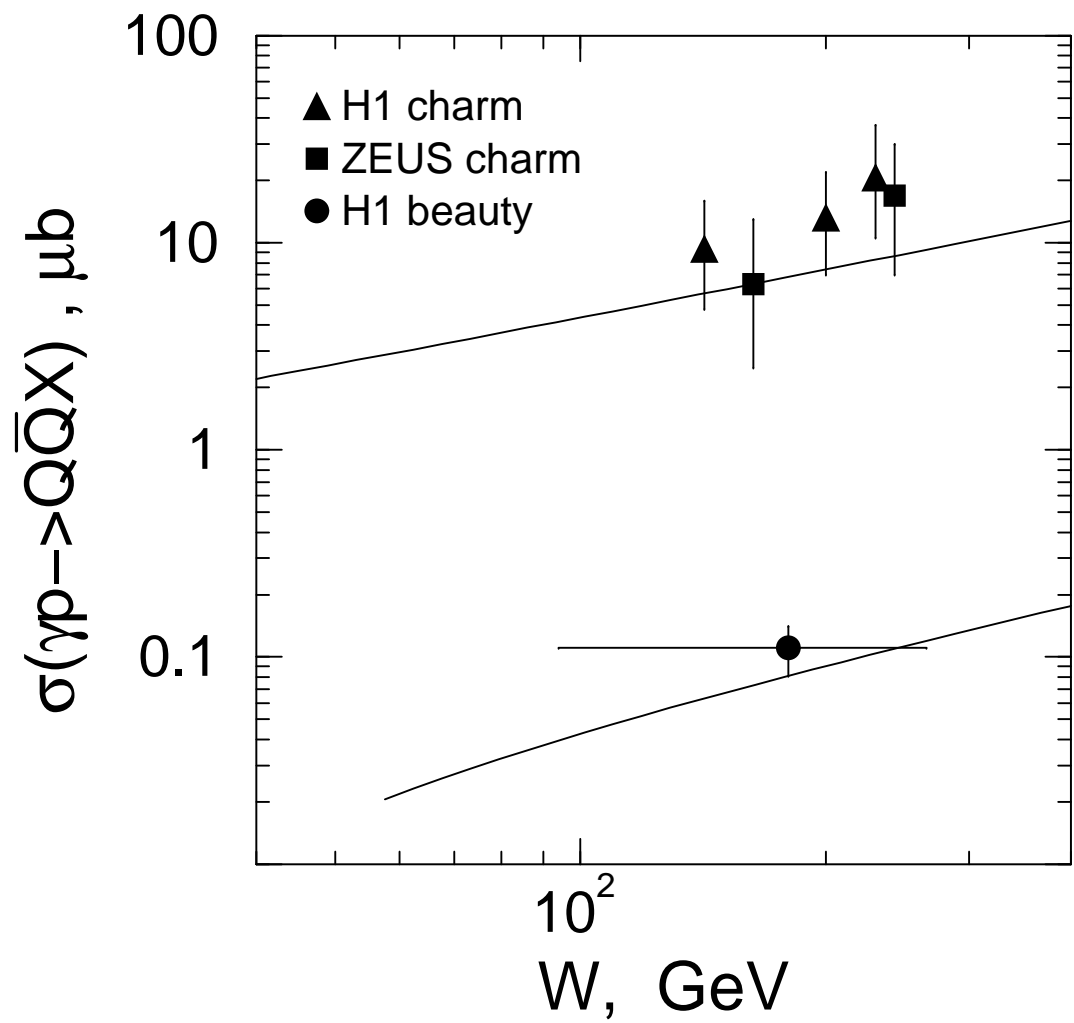
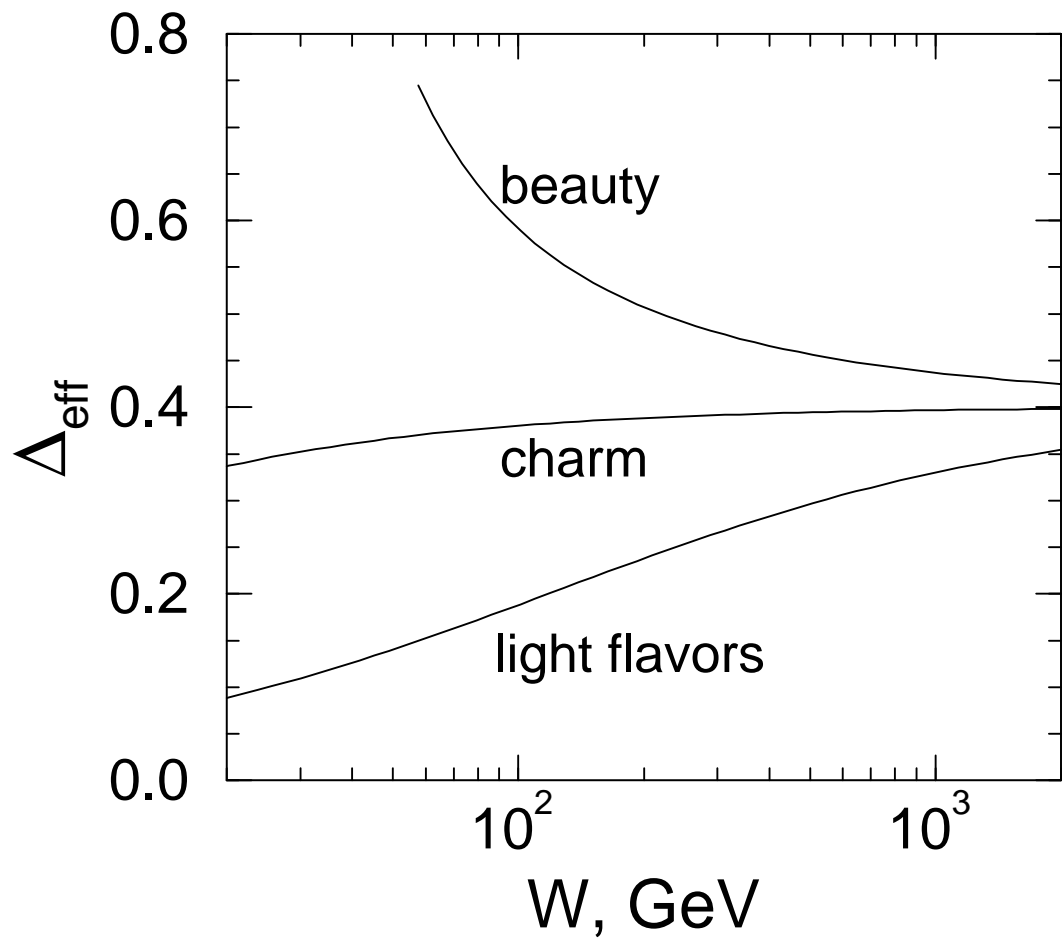
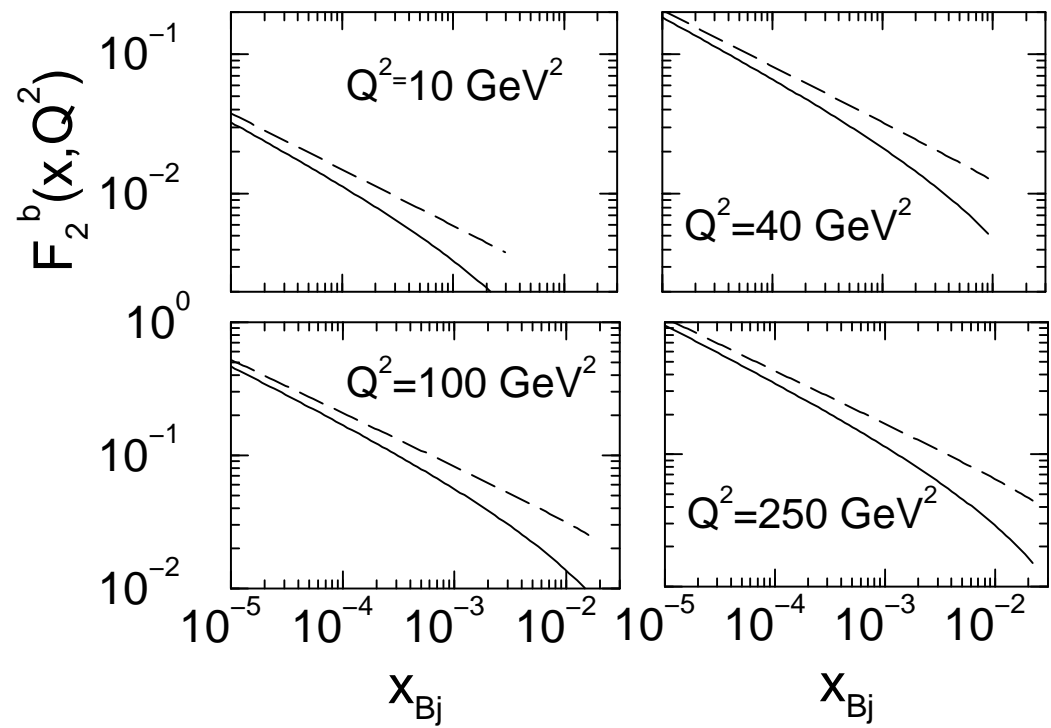
\end{document}